\documentstyle[prl,aps,multicol,epsf]{revtex}
\begin{document} 
\draft

\title{ Universal Linear Density of States for Tunneling into the 
Two-Dimensional Electron Gas in a Magnetic field }

\author{H. B. Chan, P. I. Glicofridis and R. C. Ashoori}
\address{Department of Physics, Massachusetts Institute of
Technology, Cambridge, Massachusetts, 02139}
\author{M. R. Melloch}
\address{Department of Electrical Engineering, Purdue
University, West Lafayette, Indiana  47907}
\maketitle
\date{Received}

\begin{abstract}
A new technique permits high fidelity measurement of the
tunneling density of states (TDOS) of the two-dimensional
electron gas. The obtained TDOS contains no distortions arising
from low 2D in-plane conductivity and includes the
contribution from localized tunneling sites. In a perpendicular
magnetic field, a pseudogap develops in the TDOS at the
Fermi level. Improved sensitivity enables resolution of a linear
dependence of the TDOS on energy near the Fermi energy. The
slopes of this linear gap are strongly field dependent. The data
are suggestive of a new model of the gap at low energies.
\end{abstract}


\begin{multicols}{2}
\narrowtext
Characteristics of electrons tunneling into and out of a two 
dimensional (2D) system differ considerably from those of the
three dimensional (3D) case. The distinction is especially 
pronounced when a magnetic field is applied in the tunneling 
direction (perpendicular to the plane in the 2D case). In the 
simplest picture, such a magnetic field effectively localizes 
electrons in the 2D system. An electron tunneling into an 
energetically unfavorable position cannot readily move away 
and instead tends to move in circles. As a result, tunneling 
measurements of 2D systems in a magnetic field display effects 
attributable to a ``pseudogap" in the tunneling density of 
states (TDOS) at low injection energies 
\cite{Ashoori,Eisenstein,Brown}. In contrast, for a 
3D system the tunneling electron can move parallel to the field 
lines to evade being localized at a position of high potential 
energy and such a field-induced gap has not been detected. 

There have been two chief obstacles in interpreting 
tunneling measurements of a 2D electron system. 
First, the in-plane conductance must be kept much larger than 
the tunneling conductance, otherwise the measurement 
represents transport within the 2D plane instead of across the 
tunnel barrier. This is a major problem when the 2D electron 
density is low or when the 2D electron system acts effectively 
as an insulator in the quantum Hall regime. Second, schemes 
such as tunneling between two 2D layers \cite{Eisenstein,Brown} 
yield a convolution of effects from both layers. 
Tunneling from a 2D into a spectroscopically featureless 3D 
layer achieves superior resolution of 2D features \cite{Dynes}. 
However, until now it was only possible to measure zero-bias
tunneling between 2D and 3D in semiconductor heterostructures
\cite{Ashoori}.
 
In this letter, we report results from a new technique which we 
call ``time-domain capacitance spectroscopy" (TDCS) for 
measuring the I-V characteristics of structures to which direct 
ohmic contact is not possible. Using TDCS, we measure the 
tunneling current into a 2D electron system at arbitrarily low 
values of in-plane conductivity. Unlike other tunneling current 
measurements, our technique detects {\it all} of the current 
entering or exiting the 2D layer, including those arising from 
electrons entering localized sites. 

Zero bias suppressions in tunneling conductance are well known 
to be a many body phenomenon. Different approaches have 
been taken to solve this complex problem involving 
electron-electron interactions, disorder and magnetic field 
\cite{Aleiner,He,Johansson,Pikus,Varma,Yang}. Our result 
indicates that this field-induced tunneling suppression differs 
qualitatively from previous theoretical predictions. First, the 
TDOS is found to have a {\it universal} linear dependence on 
energy near the Fermi energy for all field strengths and electron 
densities. The slopes of this linear gap are strongly field 
dependent. Second, the high excitation tunneling spectrum 
shows a change in curvature as the field strength is increased.

Figure 1a depicts the type of samples used in our experiment. 
The 2D electron system is sandwiched between two electrodes, 
close enough only to the bottom electrode to permit tunneling of 
electrons. Mesas etched from two wafers grown using 
molecular beam epitaxy have been studied. Both wafers consist 
of a degenerately n doped GaAs substrate followed by an 
AlGaAs tunnel barrier. On top of that a GaAs quantum well is 
grown which defines the 2D electron system. A thick AlGaAs 
blocking barrier prevents charge transfer between the well and 
the top GaAs electrode. The blocking barrier contains an n 
doped region to provide electrons for the well. The first wafer 
(wafer {\bf A}) \cite{Wright} has been studied in detail previously 
using frequency dependent capacitance measurements 
\cite{Ashoori} to determine zero-bias tunneling characteristics. 
The second wafer (wafer {\bf B}) \cite{Melloch} has a smaller 
dopant concentration and a thinner tunnel barrier. A DC bias 
applied 
\begin{figure}
\epsfxsize=\linewidth
\epsfbox{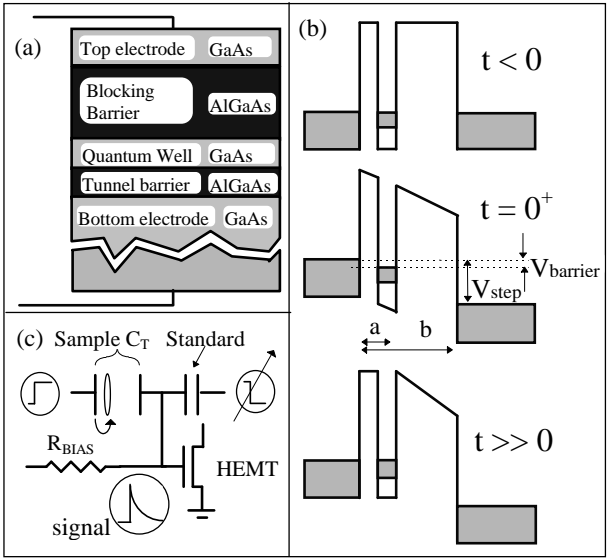}
\caption{(a) Structure of our MBE grown samples. (b) Evolution of the 
conduction band profile of our sample in one measurement cycle. 
(c) External circuit including a standard capacitor and a HEMT 
used to measure the current flowing out of the 
sample.}
\end{figure}
\noindent to the top gate permits variation of the density of 
electrons in the quantum well from depletion to 
$6 \times 10^{11} cm^{-2}$ (sample {\bf A}) and 
$3 \times 10^{11} cm^{-2}$ (sample {\bf B}) . The mobility of the 2D 
electron system in sample {\bf A} was estimated \cite{Thesis} to be 
$1 \times 10^{5}cm^{2}$/Vs at a density of 
$2 \times 10^{11} cm^{-2}$, and the mobility of sample {\bf B} is 
expected to be considerably higher. TDCS enables us to measure 
away from zero bias and extract the complete I-V characteristics. 
All features of the tunneling suppression described in this paper 
have been observed in both samples.

Figure 1b shows the evolution of the conduction band diagram 
for our samples during one cycle of the TDCS measurement. 
We start with the 2D electron system in equilibrium with the 3D 
substrate. At time t = 0, a sharp ($<10ns$ step rise) voltage step is 
applied. This creates an offset in the chemical potential on the 
two sides of the tunnel barrier, inducing a tunneling current. As 
electrons tunnel, this offset equilibrates, and the decay signal is 
recorded in real time. To measure the current across the tunnel 
barrier, a capacitance bridge is utilized (Fig. 1c). Voltage steps 
of opposite polarity are applied to the substrate of the sample 
($C_{T}$) and to one plate of a standard capacitor ($C_{S}$). The 
other plate of $C_{S}$ and the top electrode of $C_{T}$ are 
electrically connected, and the signal at this balance point is 
fed into the gate of a high electron mobility transistor (HEMT).  
Through a 70M$\Omega$ resistor $R_{BIAS}$ the DC bias of the HEMT
is established. The amplitude of the step applied to $C_{S}$ is 
adjusted so that the voltage at the balance point decays to zero 
after electrons cease tunneling. All measurements described in this 
letter take place at a temperature of 65 mK.  

Before the voltage step is applied, the 2D plane has the same 
electrochemical potential at all positions. Immediately after the 
voltage step is applied ($t=0^{+}$), no charge has been 
transferred into the 2D electron system. The voltage pulse is not 
screened by the quantum well and the 2D plane remains an 
equipotential. The simple planar geometry of the sample 
dictates that this remains true even in situations of very low 2D 
conductivity. At $t=0^{+}$, the voltage across the tunnel 
barrier ($V_{barrier}$) is simply a fraction of the voltage step 
applied ($V_{step}$), given by $V_{barrier}=(a/b)V_{step}$,
with $a$ and $b$ defined in Fig. 1b. 
The 3D substrate is always highly conducting, and electrons can 
tunnel everywhere into the 2D electron system.

Using charge conservation we have determined that there is a 
fixed relationship between the barrier current and the 
displacement (measured) current. Immediately after the applied 
voltage step, the current across the barrier is found to be 
proportional to the initial time derivative of the voltage at the 
balance point:
\begin{displaymath}
I_{barrier}=C_{\Sigma} \frac{dV_{b}}{dt} \, $ where $ 
\, C_{\Sigma}= \frac{C_{1} C_{2} - C_{1} C_{S} - C_{2} C_{S}}{C_{1}} \, 
$ (1)$
\end{displaymath}
The capacitances $C_{1}$ and $C_{2}$ in equation (1) are the 
simple geometrical values $C_{1}=\epsilon A/(b-a)$ and 
$C_{2}=\epsilon A/a$, where $\epsilon$ is the dielectric 
constant and A is the area of the mesa. Elsewhere \cite{next}, 
we prove that the relationship (1) remains true independent of 
thermodynamic DOS variations in the quantum well. By 
applying voltage steps of different amplitude and taking initial 
time derivatives of the corresponding transistor signal, the 
complete I-V characteristics of the tunnel barrier can be 
mapped out. Signals from the experiment are extremely faint, 
and immense signal averaging is involved in our measurements. 
Each point on an I-V trace may require averaging of as many as 
100,000 time traces. A novel signal processing and rapid 
averaging system \cite{patent} permits data acquisition with 
enormous ($\sim$18 bit) digital resolution. 

Figure 2 shows the tunneling conductance (I/V) of sample {\bf A}
plotted against the voltage across the barrier for magnetic field
strengths of 0, 1, 2, 8 and 16 Tesla at a fixed electron density of
$1.9 \times 10^{11} cm^{-2}$. This density is high enough so 
that no zero-bias tunneling suppression is observed at zero field.
Application of a magnetic field reduces the tunneling
conductance around zero bias. The suppression becomes 
deeper and wider as the field is increased. This field-induced 
tunneling suppression differs quantitatively from the 
logarithmic suppression \cite{Altshuler} in the low density regime at 
zero field \cite{next}. Near zero bias, the conductance is found to 
have a universal linear dependence on the excitation voltage for 
all magnetic field strengths and electron densities. Moreover, an
increase in the strength of the suppression is accompanied by a 
change in the curvature of the high excitation part of the 
conductance curves when the magnetic field is increased,
as shown by the bottom inset of Fig. 2. Even though the 
conductance curves at high excitation appear 
\begin{figure}
\epsfxsize=\linewidth
\epsfbox{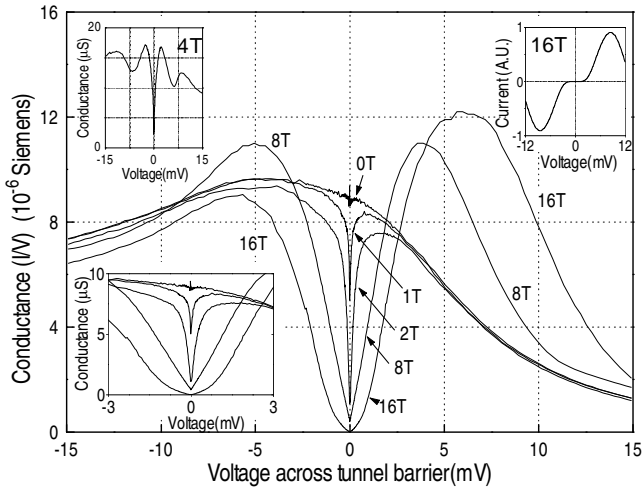}
\caption{ Dependence of the tunneling conductance (I/V) of sample {\bf A} 
on excitation voltage across the tunnel barrier for different 
magnetic field strengths at 65 mK and a fixed density of 
$1.9 \times 10^{11}cm^{-2}$. 
Bottom inset: same set of conductance curves, zooming in near zero 
bias. Top left inset: Conductance curve for sample {\bf B} at 4T and 
$\nu \sim 3$, showing Landau levels at higher excitations in addition 
to the zero bias suppression. Top right inset: Simulated I-V for 
tunneling between two 2DEG's, using 16T data from sample {\bf A} (see text).}
\end{figure}
\noindent rounded at high 
fields, the zero bias region remarkably remains linear in 
voltage, with both the magnitude and the slope significantly 
reduced. This singular behavior is illustrated by the insets of 
Fig. 3, which zoom in near the zero bias region of the 
conductance curves for different field strengths. Such a linear
energy dependence of the TDOS is observed over the full
range of densities in both samples except near depletion 
($n \leq 5 \times 10^{10} cm^{-2}$). The top left 
inset of Fig. 2 displays a conductance curve from sample {\bf B}
at a field of 4T and $\nu \sim 3$. In addition to the zero bias 
suppression, features associated with adjacent Landau levels can 
be identified at higher excitations.

In order to compare our data to results from double 
well experiments \cite{Eisenstein,Brown}, we compute the I-V 
curves expected for tunneling between two 2D electron systems: 
\begin{displaymath}
I \propto \int_{0}^{eV} g(E-eV)g(E)dE
\end{displaymath}
Here both 2D systems are assumed to have identical TDOS
$g(E)$ deduced from our 2D-3D tunneling data 
from sample {\bf A} at 16 Tesla. The resulting I-V curve, as shown in 
the top right inset of Fig. 2, qualitatively resembles that from 
double well experiments. 

To our knowledge, no existing model other than the 2D 
Coulomb gap \cite{Pikus,Efros} predicts such a universal linear 
DOS at low energies for such a wide range of field strengths. 
However, contrary to expectations for a simple Coulomb gap the
slopes of the observed linear gap are strongly field dependent. 
Figure 3 shows the slopes of the linear regions of the 
conductance curves plotted against inverse 
\begin{figure}
\epsfxsize=\linewidth
\epsfbox{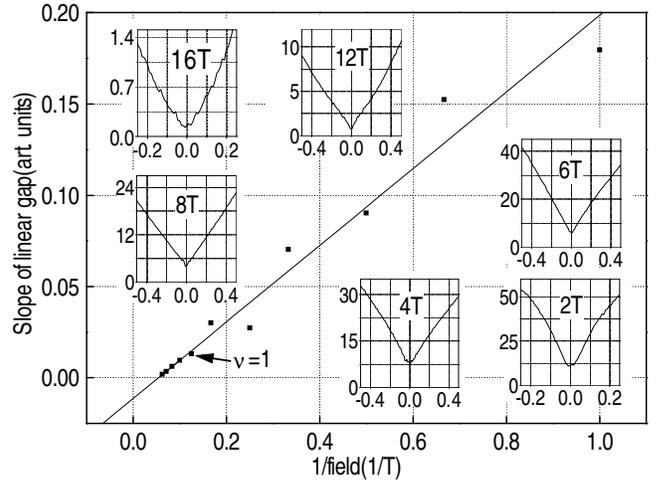}
\caption{ Insets: Tunneling conductance of sample {\bf A} as a function 
of excitation voltage for 6 different magnetic field strengths. The 
units are the same in all insets, with the abscissae in mV and the ordinates 
in $\mu$-Siemens. Different ranges are chosen to display the singular 
behavior near zero bias. Main figure: Dependence of 
the slope of this linear gap on inverse field strength.}
\end{figure}
\noindent magnetic field 
strength. For filling factors $\nu<1$, the data points fall nicely 
onto a straight line extrapolating to a negative intercept on the 
vertical axis. For low fields, there are deviations from the 
straight line as the filling fraction varies between integer and 
non-integer values.

In the Coulomb gap picture, the states in the vicinity of the 
Fermi level are assumed to be localized. These electrons are 
treated as classical point charges at fixed positions in space with 
no overlap of the electronic wavefunctions. The phase space 
available for electron tunneling is reduced since it costs more 
energy to add an electron to the system when another electron is 
located close to the tunneling electron. The resulting 
Coulombgap in the TDOS varies linearly with excitation energy 
in 2D with a slope determined solely by physical constants such 
as the electronic charge and dielectric constant \cite{Efros}. 
In contrast the slope of the linear gap in our data depends inversely 
on field strength with an offset, suggesting that a simple Coulomb
gap is inadequate in explaining the tunneling suppression in our 
experiment. 

We propose a phenomenological model which describes the 
data quite well. This model is inspired by one previously 
developed for tunneling into a system of random sized metal 
particles \cite{Cavicchi}. In this picture, the 2D system is 
modeled as isolated puddles with uniform charging energies and 
random background offsets. Interactions among the puddles are 
neglected, in contrast to the Coulomb gap model. This 
assumption may be justified due to the presence of the nearby 
3D conducting substrate which screens the interactions among 
the puddles.

\begin{figure}
\epsfxsize=\linewidth
\epsfbox{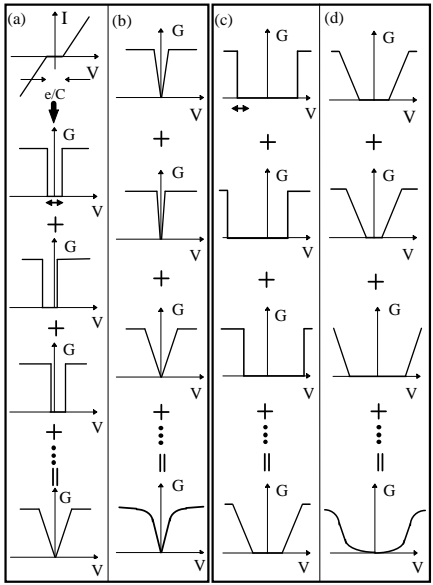}
\caption{(a) From the top: I-V curve and corresponding 
conductance curve of a single puddle. A random offset voltage 
shifts conductance curves of different puddles along the voltage 
axis. The bottom 
figure shows the V-shaped conductance curve resulting from 
summations of these randomly shifted curves. 
(b) Puddles of different sizes produce conductance curves with 
different slopes, giving rise to an overall conductance curve 
shown at the bottom. (c) When the puddles become small enough  
that e/C becomes larger than the background offset voltage, the 
overall conductance will vanish near zero bias. (d) In this 
regime, when puddles of different sizes are taken into account, 
a U-shaped conductance curve will be obtained. }
\end{figure}
Even with no applied voltage, the Fermi level of each puddle 
will not, in general, align with the Fermi level of the 3D 
bottom electrode. The energy of a puddle with capacitance C 
increases or decreases by $e^{2}/C$ when an electron is added 
to or removed from the puddle. Thus the Fermi energies on the 
two sides of the tunnel barrier are only equilibrated to 
within $e^{2}/2C$ of each other. The Fermi energies in these 
puddles are assumed to be uniformly distributed within this 
range because of, for instance, a random background voltage offset 
created by nearby dopants or impurities. Each puddle 
contributes a Coulomb blockade type I-V characteristic, leading 
to a conductance curve which is constant in voltage except for 
a region of width $e/C$ randomly displaced from zero bias where 
the conductance vanishes (Fig. 4a).

The sum of conductances from all puddles will thus be linear in 
voltage near zero bias. The slope of this linear gap is inversely 
proportional to the capacitance C, while the width of the gap is 
proportional to C. Since the capacitance C is proportional to the 
area of the puddles, our data can be explained if the average 
area of the puddles varies inversely with magnetic field strength. 
This model assumes that the high voltage conductance of an 
individual puddle is proportional to its area and the total area 
occupied by puddles is constant.

Another appealing feature of this ``Coulomb blockade gap'' model 
is that it can explain the different curvatures of the tunneling 
spectrum at high excitations as well as the negative offset of 
the slope vs. inverse field strength dependence shown in Fig. 3. 
In the low field limit, the range of random energy offsets is 
larger than the Coulomb blockade energy. When puddles of different 
sizes are taken into account, the resultant TDOS will be a 
superposition of linear gaps with different widths and slopes, 
giving rise to a negative (convex) curvature at high excitations 
while preserving the linear behavior at zero bias (Fig. 4b).
  
In the high field limit, some of the puddles become small enough 
so that their Coulomb blockade energies exceed the range of the 
background offset energy. Their conductance curve will no longer 
be V-shaped. In this regime, conductance contributions from 
puddles of a particular size will be zero at low bias up to a 
certain voltage beyond which the overall conductance rises 
linearly with voltage to the unsuppressed value, as depicted in 
Fig. 4c. Summing contributions for puddles of various sizes 
leads to the U-shaped conductance curve in Fig. 4d, concave with 
respect to voltage at high biases. As long as there exist some 
puddles large enough to produce a V-shaped conductance curve, the linear 
behavior of the overall conductance is preserved near zero bias, 
albeit with a much reduced slope. This argument can be carried 
further to explain the finite magnetic field required to produce 
a zero slope in the TDOS as extrapolated from our TDCS data. This 
happens when the puddles are small enough so that the Coulomb 
blockade energy of every puddle in the system exceeds the range 
of the background offset energies. It is not necessary to have an 
infinitesimal puddle to achieve a zero slope for conductance near 
zero bias. 

The above deductions are based on the assumption that 
larger puddles break up into smaller ones and that the mean area 
of the puddles shrinks linearly with increasing field strength. 
Electrons therefore charge parallel plate capacitors whose 
lateral dimension is proportional to the magnetic length. From 
the width of the gap in our data, we estimate the proportionality 
constant to be about 6. We do not yet have a clear answer to 
the question of what the puddles are or why they shrink as the 
magnetic field strength is increased.

While this simple picture of a Coulomb blockade gap may not provide 
a complete description of the system, it seems to be able to explain 
qualitatively most features in our data. A more thorough understanding 
of the tunneling suppression will require inclusion of interaction, 
charging effects and the quantum mechanical properties of the 2D 
electron system. 

We thank L. S. Levitov, A. V. Shytov, J. K. Jain, R. K. Kamilla 
and A. H. MacDonald for helpful discussions and M. Brodsky 
for assistance in the experiment. This work is supported by the 
ONR, JSEP-DAAH04-95-1-0038, the Packard Foundation, 
NSF DMR-9357226, DMR-9311825, and DMR-9400415.

\end{multicols}

\end{document}